\documentclass{article}

\usepackage{arxiv}
\usepackage{hyperref}
\hypersetup{
  colorlinks=true,
  linkcolor=blue,      
  urlcolor=blue,
  citecolor=blue
}
\usepackage{url}
\usepackage{amsmath}
\usepackage[fleqn,tbtags]{mathtools}
\usepackage[numbers]{natbib}
\usepackage{acronym}
\usepackage{csquotes}
\usepackage{array}
\usepackage{booktabs}
\usepackage{enumitem}

\title{Solving Distributed Flexible Job Shop Scheduling Problems in the Wool Textile Industry with Quantum Annealing}

\author{
 Lilia Toma\\
  Databases and Information Systems\\
  FernUniversität in Hagen\\
  58097 Hagen, Germany\\
  \texttt{tomalilia@yahoo.com}\\
   \And
 Markus Zajac\\
  Databases and Information Systems\\
  FernUniversität in Hagen\\
  58097 Hagen, Germany\\
  \texttt{markus.zajac@fernuni-hagen.de}\\
  \And
 Uta St{\"o}rl\\
  Databases and Information Systems\\
  FernUniversität in Hagen\\
  58097 Hagen, Germany \\
  \texttt{uta.stoerl@fernuni-hagen.de}\\
}

\begin{document}
\maketitle

\newacro{BQM}{Binary Quadratic Model}
\newacro{CPU}{Central Processing Unit}
\newacro{DFJSP}{Distributed Flexible Job Shop Scheduling Problem}
\newacro{DBC}{D-Wave Binary CSP}  
\newacro{FJSP}{Flexible Job Shop Scheduling Problem}
\newacro{GA}{Genetic Algorithm}
\newacro{JSSP}{Job Shop Scheduling Problem}
\newacro{FSP}{Flow Shop Scheduling Problem}
\newacro{NISQ}{Noisy Intermediate-Scale Quantum}
\newacro{SA}{Simulated Annealing}
\newacro{SAPI}{Solver API}
\newacro{SDK}{Software Development Kit}
\newacro{QA}{Quantum Annealing}
\newacro{QUBO}{Quadratic Unconstrained Binary Optimization}
\newacro{QPU}{Quantum Processing Unit}

\begin{abstract}
  Many modern manufacturing companies have evolved from a single production facility to a multi-factory production environment that must manage both regionally dispersed production orders and their multi-site production steps. The availability of a range of machines in different locations capable of performing the same operation and shipping times between factories have transformed planning systems from the classic \ac{JSSP} to the \ac{DFJSP}. Consequently, the complexity of production planning has increased significantly. We employ \ac{QA} to solve the \ac{DFJSP} in our research. In addition to assigning production orders to production sites, production steps are also assigned to these sites. This requirement is based on a real use case of a wool textile manufacturing company. To investigate the applicability of this method to large problem instances, problems ranging from 50~variables up to 250~variables, the largest problem that could be embedded into a D-Wave quantum annealer \ac{QPU}, are formulated and solved. Special attention is dedicated to determining the Lagrange parameters of the \ac{QUBO} model and the \ac{QPU} configuration parameters, as these factors can significantly impact solution quality. The obtained solutions are compared to solutions obtained by \ac{SA}, both in terms of solution quality and calculation time. The results demonstrate that \ac{QA} has the potential to solve large problem instances specific to the industry.
\end{abstract}


\section{Introduction}\label{sec:intro}

This paper is based on a use case of a textile manufacturing company that operates in the worsted yarn industry, specializing in pure wool and wool blends for weaving, flat and circular knitting, as well as luxury and non-apparel markets. This business-to-business company has production sites in several countries on different continents. Production sites in different locations perform only specific production steps and work together in sequence-dependent setups to fulfill each individual sales order. For this purpose, sales orders are first converted into production orders, which must be assigned to the various production sites for processing through different production steps. Given the scattering of production sites that may be involved in processing a single production order, thorough production planning is crucial.

The reason is that this production planning problem is an NP-hard combinatorial optimization problem, which can be reduced to the Distributed Flexible Job Shop Scheduling Problem (\ac{DFJSP}) in this use case. It is widely known as one of the most difficult combinatorial optimization problems~\cite{110:Marzouki2018}. At the same time, operational production planning requires continuous and process-parallel optimization of machine allocation and processing sequence.

Quantum annealing (\ac{QA}) is a promising technology to solve combinatorial optimization problems. While gate-based quantum computers can handle a broader range of problems, quantum annealers are specifically designed to solve combinatorial optimization problems commonly encountered in industry~\cite{149:Yarkoni2022}. In this work, we use the \textit{Advantage System 4.1} quantum annealer from D-Wave to solve the \ac{DFJSP}. 

To the best of our knowledge, there is no other work that explicitly addresses solving the \ac{DFJSP} using \ac{QA}. Our three contributions are as follows:

\begin{enumerate}
    \item We use a \ac{QA} to solve the \ac{DFJSP}. Due to the specifics of the wool textile industry, we also extend the \ac{DFJSP} and make it even more complex:~The classical \ac{DFJSP} is extended to not only consider the distribution of production orders, but also the distribution of production steps (of a production order). Shipping times between production steps at different sites must also be included in the model. To the best of our knowledge, there is no other work in the literature that discusses the solution of the classical or the extended \ac{DFJSP} with \ac{QA}.
     
    \item Calculation of \textit{Lagrange parameters} and \ac{QPU} configuration parameters. The values of the so-called Lagrange parameters of the \ac{QUBO} model and the \ac{QPU} configuration parameters have a significant influence on the solution quality. Instead of the trial-and-error method, as some approaches suggest\footnote{Programming the D-Wave QPU [White Paper]:~\url{https://www.dwavesys.com/media/qvbjrzgg/guide-2.pdf}}, we determine these parameters based on the mathematical formulation of the problem.    

    \item Assessment from an economic perspective. We address the question of whether the use of \ac{QA} can lead to a speed advantage for the \ac{DFJSP} with distributed operations specific to the wool textile industry.
\end{enumerate}

This paper is organized as follows: In Section~\ref{sec:fundamentals}, we review the state of the art on optimization problems and quantum annealing, before we discuss related work in Section~\ref{sec:work}. We describe our QUBO model to solve the extended \ac{DFJSP} in Section~\ref{sec:modell}. Based on the QUBO model, we describe and discuss our experiments in Section~\ref{sec:exp}. Section~\ref{sec:conclusion} provides a summary.
\section{Preliminaries}\label{sec:fundamentals}
In Subsection~\ref{sec:fundamentals:comb}, we initially study computationally complex combinatorial optimization problems that are typical in production planning. We then turn to the topic of quantum annealing in Subsection~\ref{sec:fundamentals:qa} and to simulated annealing in Subsection~\ref{sec:fundamentals:sa}. The connection between the \ac{BQM}~/~\acl{QUBO} (\ac{QUBO}) and \ac{QA} is highlighted in Subsection~\ref{sec:fundamentals:bqm}.

\subsection{Combinatorial Optimization Problems}\label{sec:fundamentals:comb}

The \acl{JSSP} (\ac{JSSP}) is a combinatorial optimization problem\footnote{Combinatorial optimization is the search process for maxima (or minima) of an objective function where the domain has a discrete but large configuration space.} where a given set of jobs (or production orders) must be executed. At the same time, only a limited number of machines can perform the operations required for the jobs (also referred to as production steps). Each operation can be executed by a specific machine. A schedule must be found that determines which jobs are to be executed at which time and by which machine. The aim is to minimize the total amount of time until all jobs have been completed~\cite{102:Carugno2022, 156:Baskar2012}. Many variations of the \ac{JSSP} problem exist. To be closer to both the market and manufacturing resources, modern enterprises have evolved from a single factory to a multi-factory manufacturing environment. This has led to the scheduling system changing from a \ac{JSSP} system to a \ac{FJSP} and further to a \acl{DFJSP} (\ac{DFJSP}).

The \ac{FJSP} is a generalization of the \ac{JSSP} that allows tasks to be processed on one machine, chosen from a set of alternative machines (the machines in each set are identical)~\cite{200:Nouri2017}. As is well known, this is a strongly NP-hard problem~\cite{139:Guohui2019}. A distinctive feature of \ac{JSSP} (and \ac{FJSP}) is the parallel job processing. While different process chains for an order are evaluated, the selection of a process chain for an order influences the choice of production routes for all other orders.

The \ac{DFJSP} refers to production activities that are carried out by several geographically distributed factories. The \ac{DFJSP} can be regarded as an extension of \ac{FJSP}, which also includes the selection of appropriate factories for the jobs~\cite{114:Liu2015}. The \ac{DFJSP} described in the literature implies that if a job is assigned to a factory, then all the operations of this job must be processed at the same factory~\cite{114:Liu2015,110:Marzouki2018}. The wool textile industry is a special case of \ac{DFJSP}, where a job's operation can be processed in different locations. Assigning specific operations to different factories results in unique production schedules for the jobs, which influences the production planning. The \ac{DFJSP} is more complex than \ac{FJSP} and is also classified as NP-Hard in complexity theory~\cite{110:Marzouki2018}. The goal of scheduling is to assign each operation to factories and machines, considering the shipping time between locations, and launching operations on machines in a way that minimizes the makespan of all operations, thus ensuring that all jobs are completed in the shortest time possible.

\subsection{Quantum Annealing}\label{sec:fundamentals:qa}

Today, two main architectures for quantum computers exist: \mbox{gate-based} quantum computers and \ac{QA}~\cite{102:Carugno2022}. \mbox{Gate-based} quantum computers use quantum gates to perform calculations, analogous to Boolean gates in classical computers. These processors are not yet fault-tolerant or large enough to achieve quantum supremacy, a state referred to as the \ac{NISQ} era~\cite{152:Brooks}. Consequently, it is difficult to develop practical applications on \mbox{gate-based} quantum computers at present.

On the other hand, \ac{QA} devices, such as the computer developed by D-Wave, map a problem to a mathematical model of interacting magnets and then search for an optimal solution for problems with a large solution space. \citeauthor{Liu2024}~\cite{Liu2024} developed a large-scale and energy-efficient superparamagnetic Ising quantum annealer, which advances the scope of coherent Ising devices for combinatorial optimization. In general, \ac{QA} is a mechanical metaheuristic to solve problems with a large, but discrete, search space.

\ac{QA} systems are initialized in a low-energy state. The problem parameters are changed slowly enough, which allows the system to end in a low-energy state of the problem that corresponds to an optimal solution~\citep{149:Yarkoni2022}. The evolution of the system can be represented by a time-dependent Hamiltonian\footnote{See also~\citeauthor{102:Carugno2022}~\citep{102:Carugno2022} and \url{https://docs.dwavequantum.com/en/latest/quantum_research/quantum_annealing_intro.html} for the Hamiltonian term.}:

\begin{equation}
\mathcal{H}_{i s i n g} =\underbrace{-\frac{A(s)}{2}\left(\sum_i \hat{\sigma}_x^{(i)}\right)}_{\text {Initial Hamiltonian }} +\underbrace{\frac{B(s)}{2}\left(\sum_i h_i \hat{\sigma}_z^{(i)}+\sum_{i>j} J_{i, j} \hat{\sigma}_z^{(i)} \hat{\sigma}_z^{(j)}\right)}_{\text {Final Hamiltonian }}
\end{equation}\label{eq:hamiltonian}

where \emph{A} and \emph{B} are time-dependent coefficients, \emph{s} is the annealing time rescaled to take values between \emph{0} and \emph{1}, $\hat{\sigma}_{x, z}^{(i)}$ are the Pauli Matrices\footnote{In quantum mechanics, Pauli Matrices are a set of 2 × 2 complex Hermitian and unitary matrices that arise in Pauli's treatment of spins.} of operations on a qubit \emph{q$_{i}$}, \emph{h$_{i}$} is the qubit bias, and \emph{J$_{i,j}$} the coupling strength between qubits. The algorithm gradually transitions in time from the initial ground state of the \emph{Initial Hamiltonian} to a state described by the \emph{Final Hamiltonian}~\citep{103:Schworm2023}. The {Final Hamiltonian} matches the energy function to be minimized, and represents the best solution at its minimum value. 

\subsection{Simulated Annealing}\label{sec:fundamentals:sa}

\begin{figure}[ht]
    \centering
    \includegraphics[width=0.8\textwidth]{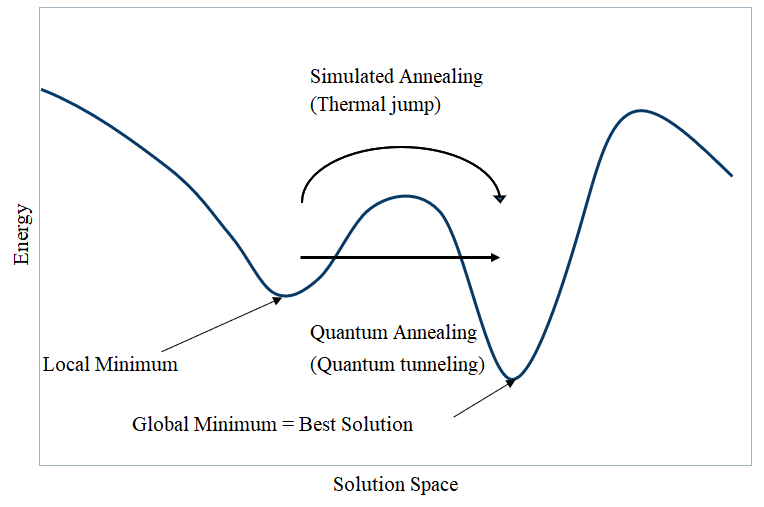}
    \caption{Quantum Annealing versus Simulated Annealing (Source: authors, based on~\citeauthor{157:Kasirajan2021}~\citep{157:Kasirajan2021}).}
    \label{fig:subsec:background:qasa}
\end{figure}

\acl{SA} (\ac{SA}) is a nature-inspired stochastic metaheuristic, single solution-based algorithm used to find a global optimum in a large problem space~\citep{Talbi2009}, and can be seen
as a classical computing pendant to \ac{QA}. Fig.~\ref{fig:subsec:background:qasa} shows a comparison between \ac{QA} and \ac{SA}.

In \ac{SA}, to simulate the annealing process, a temperature variable is set high, and then the system is allowed to cool while the algorithm runs. \ac{SA} is using thermal effects to converge to a low energy state, but therefore needs to jump over higher energy barriers. On the other hand, \ac{QA} uses the quantum tunneling effect\footnote{In quantum mechanics, the tunneling effect describes the penetration of particles through barriers, even if the particle's energy is less than the barrier height.}, which allows it to potentially surpass this higher barrier directly and thus reach the minimum faster~\citep{129:Denchev2016}. Recent advancements in \ac{QA} methods have suggested potential benefits by employing \ac{QA} over \ac{SA} to address NP-hard optimization challenges~\citep{Mukherjee_2015}.

\subsection{Binary Quadratic Models}\label{sec:fundamentals:bqm}

For \ac{QA}, cost functions are mathematical expressions that are usually formulated as binary quadratic models\footnote{\url{https://docs.dwavequantum.com/en/latest/quantum_research/qubo_ising.html}} and whose lowest values generally represent the best solutions to problems~\citep{149:Yarkoni2022,121:zaman2021pyqubo}. To solve problems on the D-Wave quantum annealer, the problem space has to be formulated as a \ac{BQM}, the D-Wave Systems software representation of the objective function~\citep{121:zaman2021pyqubo}:
 
\begin{equation}
H_{\text {Ising }}(s)=\sum_{i \in V} h_i s_i+\sum_{(i j) \in E} J_{i j} s_i s_j, \hspace{0.5em} s_i \in\{-1,1\}\label{eq:ising}
\end{equation}

\begin{equation}
H_{\mathrm{QUBO}}(\boldsymbol{x})=\sum_{i \in V} a_i x_i+\sum_{(i j) \in E} b_{i j} x_i x_j, \hspace{0.5em} x_i \in\{0,1\}\label{eq:qubo}
\end{equation}

where \emph{s$_{\text{i}}$} and \emph{x$_{\text{i}}$} are binary variables, i.e., categorical variables that can only take one of two values. \emph{s$_{\text{i}}$} is the decision variable called spin that can take values from the \mbox{set \{-1,1\}}, in which case the problem is called an Ising problem (Eq.~\ref{eq:ising})~\citep{82:Lucas2014}. In the Ising model, \emph{h$_{i}$} is the magnetic field that represents the linear coefficients corresponding to qubit biases and \emph{J$_{i,j}$} represents the coupling or interaction strength between qubits~\citep{121:zaman2021pyqubo,149:Yarkoni2022}. In a \ac{QUBO} problem (Eq.~\ref{eq:qubo}), \emph{x$_{\text{i}}$} can take values from the set \{0,1\}~\citep{96:lewis2017}. Qubit biases are represented by the linear coefficients \emph{a$_{i}$} and the coupling strength between qubits by the quadratic coefficients \emph{b$_{i,j}$}~\citep{doi:10.7566/JPSJ.88.061010}. Both versions are NP-hard and can easily be converted into each other by applying variable
substitution~\citep{doi:10.7566/JPSJ.88.061010,106:Vyskocil2019}. In this paper, the \ac{QUBO} format is used to formulate the problem. This is because QUBO formulations are a widely used method for this~\citep{DBLP:journals/anor/GloverKHD22}.

The QUBO formulation is further mapped to a QPU topology~\citep{149:Yarkoni2022} to solve the problem during the annealing process and minimize the energy function. \ac{QA} devices are less affected by noise than gate model quantum computers. This allows for more qubit usage and thus solves larger problems. \ac{QA} provides an approach that promises good solutions to NP-hard problems and specializes in combinatorial optimization. While the discussion continues whether \ac{QA} can provide quantum speed-up over classical approaches, there have been several attempts to solve application-motivated problems with \ac{QA}~\cite{149:Yarkoni2022}.

\section{Related Work}\label{sec:work}
In this Section, we assess previous approaches to solving the optimization problems described in Subsection~\ref{sec:fundamentals:comb}. 

\textbf{\ac{JSSP}.} In the past, a variety of approaches to solve the \ac{JSSP} or its variation existed. \citeauthor{68:Denkena2021}~\cite{68:Denkena2021} mention heuristics, \ac{GA}, Tabu Search, Ant Colony, Particle Swarm Optimization and \ac{SA}. These approaches are used both individually and in hybrid combinations with one another. \citeauthor{49:venturelli2016}~\cite{49:venturelli2016} present a solver for the \ac{JSSP} that makes use of a \ac{QA}. \citeauthor{102:Carugno2022}~\cite{102:Carugno2022} studied the application of \ac{QA} to solve \ac{JSSP} and discussed challenges related to problem formulation for the fine-tuning of the \ac{QA}. In terms of solution quality, \ac{QA} is remarkably effective in comparison to classical approaches on very constrained problem instances, while closely competitive on realistic problem instances. \citeauthor{120:Pakhomchik2022}~\cite{120:Pakhomchik2022} focus on the mathematical formulation of \ac{JSSP} and propose a new large neighborhood search approach, with good empirical results for large problem instances.

\citeauthor{123:Zielewski2020}~\cite{123:Zielewski2020} use \ac{QA} to solve \ac{JSSP} and especially investigate the embedding process in the performance. The authors demonstrate that the initial setup of the problem, post-processing, and problem formulation, as well as the presence of excess variables and large embeddings, have a significant impact on the results. \citeauthor{117:sharma2018}~\cite{117:sharma2018} compare \ac{QA} to \ac{SA} through existing published works and conclude that \ac{QA} has the potential to be considered as an optimized technique against \ac{SA}. \ac{QA} is more efficient than \ac{SA} in estimating the ground state of the system for simulations of small and large sample sizes. \citeauthor{129:Denchev2016}~\cite{129:Denchev2016} show that, for problems where local minima is separated by tall and narrow energy barriers, \ac{QA} is more than $10^\textsuperscript{8}$ times faster than \ac{SA} running on a single core. \citeauthor{DBLP:WeiFGGHW24}~\cite{DBLP:WeiFGGHW24} introduce a hybrid quantum-classical Benders' decomposition for federated learning scheduling, reducing iteration counts by 70.3{\%} and computation time by 81{\%}. \citeauthor{Tomasz2024}~\cite{Tomasz2024} show that a hybrid quantum-classical computation scheme can be efficiently used for automatic guided vehicle scheduling in manufacturing.

In general, research considering transportation time focuses on the flow shop scheduling problem\footnote{Flow shop scheduling is a special case of job shop scheduling in which all jobs have identical operation flow patterns on machines~\cite{156:Baskar2012}.} or classical \ac{FJSP}. Most of the \ac{QA} research in this field focuses on solving \ac{JSSP} or \ac{FJSP}, and transportation time is often neglected in the literature. However, transportation time between locations can have a significant impact on the maximum completion time of \ac{FJSP}. 

\textbf{\acs{FSP}.} \ac{FSP} is one of the most studied problems in order planning. \citeauthor{145:Hurink2005}~\cite{145:Hurink2005} apply \ac{SA} to minimize both the total completion time and the total tardiness of flow shop problems. \citeauthor{146:Naderi2009}~\cite{146:Naderi2009} propose an adaptation of the \ac{SA} algorithm to solve the flow shop problem with incorporated sequence-dependent transportation times with a single-transporter. Several heuristic algorithms were developed considering the transportation of an interrupted job from one machine to another. \citeauthor{147:Naderi2009}~\cite{147:Naderi2009} consider the transportation times in permutation flow shops and provide solution methods that include heuristics and metaheuristics to solve large problems. An adaptation of the imperialist competitive algorithm, hybridized with a simulated annealing-based local search, was proposed by~\citeauthor{148:Karimi2018}~\cite{148:Karimi2018} to solve the problem.

\textbf{\ac{FJSP}.} \citeauthor{103:Schworm2023}~\cite{103:Schworm2023} investigate a \ac{QA}-based \ac{FJSP} on large problem instances specific to modern manufacturing and demonstrate the efficiency of the approach regarding scalability, solutions' quality, and computing time. \citeauthor{68:Denkena2021}~\cite{68:Denkena2021} compare the computing time of several approaches to solve the classical \ac{FJSP} and conclude that it generally increases significantly with the size of the problem. Their research suggests that these methods deliver valid results for small problem instances, but are not suitable for problem sizes relevant to practical applications. Therefore, methods are required that yield high-quality outcomes while also ensuring consistently quick computation times, particularly for large problem instances. The paper also investigates the concept of \ac{QA}- based optimization for a process-parallel \ac{FJSP} on existing benchmarks. The results reveal that the \ac{QA} technique consistently achieves low anneal times, especially for large instances, while still producing outcomes comparable to those of classical heuristics. Recent work by \citeauthor{Fu_Liu_Chen_Zhang_2025}~\cite{Fu_Liu_Chen_Zhang_2025} solves the \ac{FJSP} using a Coherent Ising Machine (CIM) and shows an 80{\%} improvement compared to classical methods, but only for small instances, due to existing hardware constraints. \citeauthor{jstPracticalApproach}~\cite{jstPracticalApproach} analyzed the \ac{FJSP} with Tool Switching Constraints with the Constrained Quadratic Model (CQM) solver supported by the D-Wave Leap Hybrid Solver and showed that the quality and computation time significantly outperform the results achieved using Dispatching rules or Python Mixed-Integer Linear programs. \citeauthor{Schworm2023}~\cite{Schworm2023} addressed real-time responsiveness to disturbances in manufacturing environments using \ac{QA} for dynamic job shop scheduling and showed an advantage for large problem sizes. \citeauthor{Amaro2022}~\cite{Amaro2022} present a case study applying variational quantum algorithms to job shop scheduling with notably accelerated convergence. Recent results by \citeauthor{Zhang2024}~\cite{Zhang2024} show that quantum computing can deliver dramatic efficiency gains for several automated guided vehicles scheduling models in manufacturing, with up to 92{\%} time savings. \citeauthor{Chen2024}~\cite{Chen2024} introduce a dynamic \ac{FJSP} approach that incorporates machine breakdowns and new job arrivals, thereby furthering the practicality of quantum-inspired scheduling.

\textbf{\ac{DFJSP}.} Only few papers focus on the \ac{FJSP} with transportation time between jobs, also known as \ac{DFJSP}. \citeauthor{Ziaee2017DFJSP}~\cite{Ziaee2017DFJSP} developed a novel mathematical model for the distributed \ac{DFJSP} that explicitly incorporates transportation time between machines, and proposed an effective optimization framework that significantly reduces makespan and enhances coordination across distributed manufacturing environments. \citeauthor{Peng2025}~\cite{Peng2025} formulated a multi-objective dynamic \ac{DFJSP} that explicitly accounts for uncertain processing times, proposing a novel optimization framework that improves schedule robustness and adaptability in distributed manufacturing environments under real-world uncertainty. \citeauthor{YUAN2025101902}~\cite{YUAN2025101902} proposed an improved deep Q network approach for the distributed heterogeneous \ac{FJSP} that integrates automated guided vehicle (AGV) transportation, demonstrating substantial optimization of production makespan and resource utilization. \citeauthor{139:Guohui2019}~\cite{139:Guohui2019} develop a modified \acf{GA} to solve the \ac{FJSP} with transportation time. \citeauthor{110:Marzouki2018}~\cite{110:Marzouki2018} use a Chemical Reaction  Optimization Metaheuristic to solve the \ac{DFJSP}. \citeauthor{114:Liu2015}~\cite{114:Liu2015} propose a refined \ac{GA} and apply it with satisfactory results on a real case of a manufacturing company. 

Generally, most of the papers focus on studying the \ac{JSSP}. Although some research includes transportation time in the model, these studies typically employ classical approaches, such as heuristics or \ac{GA}. In this research, an extended case of \ac{DFJSP} is studied, where not only jobs (production orders), but also job operations (production steps) can be distributed. For this purpose, a D-Wave Quantum Annealer is used to solve the \ac{DFJSP} with distributed operations. The availability of a set of machines capable of processing the same operation and the geographical distribution of these machines add complexity and pose a significant challenge to solve this combinatorial optimization problem.
\section{Modell}\label{sec:modell}
In this Section, we describe the QUBO model for solving the extended \ac{DFJSP} as required by the wool textile industry. To this end, the extended problem is first formally described in Subsection~\ref{sec:problem_formulation}. Based on this, we create the QUBO model in Subsection~\ref{sec:qubo_model}, reduce the number of its variables in Subsection~\ref{sec:variable_pruning}, and deal with the determination of the Lagrange parameters in Subsection~\ref{sec:penalties}.

\subsection{Problem formulation}
\label{sec:problem_formulation}
The \ac{DFJSP} refers to job scheduling in distributed manufacturing environments. There could be different objectives when solving the \ac{DFJSP} (e.g., minimizing job completion time, processing costs, or energy consumption). This paper focuses on finding the optimal schedule that minimizes the time required to complete the jobs, i.e., the  \emph{makespan}. The formulation used is the one introduced by~\citeauthor{49:venturelli2016}~\citep{49:venturelli2016} for the \ac{JSSP} and the extension from~\citeauthor{103:Schworm2023}~\citep{103:Schworm2023} for the \ac{FJSP}. In this research, the model is extended to include shipping time for distributed operations. It is assumed that there are A jobs \emph{{J = \{j$_\emph{{\text{1}}}$, ..., j$_\emph{{\text{A}}}$\}}} to be processed on B machines \emph{{M = \{m$_\emph{{\text{1}}}$, ..., m$_\emph{{\text{B}}}$\}}} in C factories \emph{{F = \{f$_\emph{{\text{1}}}$, ..., f$_\emph{{\text{C}}}$\}}}. Each job \emph{${i\in J}$} consists of a set of L operations \mbox{\emph{{${O_{i}}$ = \{o$_\emph{{\text{1}}}$, ..., o$_\emph{{\text{L}}}$\}}}}. Any operation \emph{${o\in O_i}$} can be processed on at least one machine out of the given set \emph{M$_{\text{i,o}}$}. For any job, the order of operations \emph{${o\in O_i}$} is given by \emph{{${R_{O_{i}}}$ = \{r$_{\text{o}_\text{1}}$, ..., r$_{\text{o}_\text{l}}$\}}}, in which \emph{r$_{\text{l}}$} is the last operation of job \emph{${i}$}. Another notations used in this paper are listed in Tbl.~\ref{tab:modell_notation}.

\begin{table}[ht]
\begin{center}
%\begin{tabular}{l l}
\begin{tabular}{p{50pt}p{300pt}}
\toprule
Notation& 
Description \\
\midrule
\emph{o$'$}& 
Next operation of any operation \emph{${o\in O_i}$}. \\
\emph{m$'$}& 
Machine of next operation \emph{o$'\in O_i$}. \\
\emph{t}& 
Time step in a given timeline \emph{{T = \{0, ..., t$_\emph{{\text{max}}}$\}}}, where \emph{t$_{\text{max}}$} is the completion time of all jobs. \\
t$'$& 
Next time step of any time step \emph{${t\in T}$}. \\
\emph{t$_{\text{i,o,m}}$}& 
Starting time of operation \emph{o} of job \emph{i} on machine \emph{m}. \\
\emph{p$_{\text{i,o,m}}$}& 
Processing time of operation \emph{o} of job \emph{i} on machine \emph{m}. \\
\emph{d$_{\text{m,$m'$ }}$}& 
Shipping time between machines. \\
\emph{k$_{\text{f,$f'$ }}$}& 
Shipping time between factories. \\
%\emph{c$_{\text{i}}$}& 
%Completion time of a job. \\
\emph{P$_{\text{i,o}}$}& 
Minimum predecessor time of operation \emph{o}. \\
\emph{S$_{\text{i,o}}$}& 
Minimum successor time of operation \emph{o}. \\
\bottomrule
\end{tabular}
\end{center}
\caption{List of notations used in this paper.}
\label{tab:modell_notation}
\end{table}

\begin{figure}[ht]
    \centering
    \includegraphics[width=0.8\textwidth]{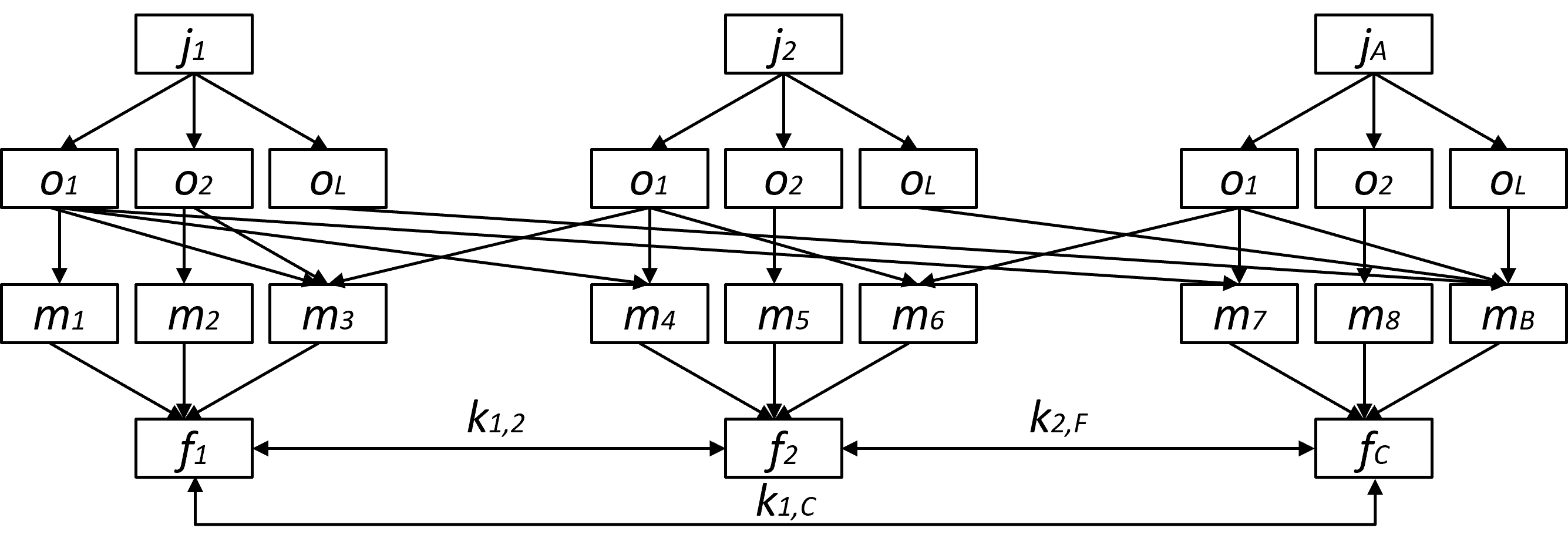}
    \caption{Presentation of the \ac{DFJSP} consisting of \textit{A} Jobs, \textit{L} Operations (per Job), \textit{B} Machines and \textit{C} Factories (Source: authors).}
    \label{fig:dfjsp}
\end{figure}

Reminder: The production orders are referred to as jobs and the production steps as operations in the context of the \ac{DFJSP}. Since machines are associated with factories, the distance between machines is derived from the distance between factories. Distance between factories \emph{$k$$_{\text{$f$,$f'$ }}$} is directly related to the shipping time \emph{$d$$_{\text{$m$,$m'$ }}$} needed to send the lots in production to another location. In this paper, 'distance' and 'shipping time' are used interchangeably and refer to the value added to the cost function when production lots are shipped between locations. Fig.~\ref{fig:dfjsp} shows a schema representing the \ac{DFJSP}.

Different jobs may have varying numbers of operations, each with its own processing time on a specific machine. This paper focuses on a special case of \ac{DFJSP} where operations of the same job can be performed on machines in geographically distributed factories, which means that the distance between machines should be considered when generating the job schedule.

In the \ac{DFJSP} the following constraints are considered: 

\begin{enumerate}
 \item Precedence of operations - an operation can only start when the previous operation has finished.
 \item Operation starts once - operation processing on machines cannot be interrupted (the reason ,behind this condition is that each machine must be technically adjusted for every new production lot, which is time-consuming).
 \item No-overlap - each machine can only process one operation at a time.
\end{enumerate}
 
Other underlying assumptions of the \ac{DFJSP} are that operations of a job must be performed in the correct order to complete that job. The operations of different jobs are not subject to any precedence constraints and can be processed in parallel. Each factory is equipped with machines of various types, resulting in different potential processing times for an operation. A job can be scheduled on a machine multiple times. All machines are available for use and all jobs can be processed starting with time zero~\citep{139:Guohui2019}.

%
% Section: QUBO model
%
\subsection{QUBO model}
\label{sec:qubo_model}
Samplers are processes used to solve problems with D-Wave. A \ac{BQM} sampler samples low-energy states in the model defined by the \ac{QUBO} equation and returns a list of samples to increase energy. The \ac{QUBO}  model is composed of binary variables \emph{x$_{\text{i,o,m,t}}$}, each representing one operation per unit of time on a machine, up to the makespan:

\begin{equation}
x_{i, o, m, t}=\left\{\begin{array}{l}
\begin{aligned}
1:~&\text{operation } o \text{ of job } i \text{ starts on }\text{machine } m \text{ at time } t\\
0:~&\text{ otherwise}
\end{aligned}
\end{array}\right.
\label{eq:bin_var2}
\end{equation}

This variable determines whether an operation \emph{o} of a job \emph{i} starts on a machine \emph{m} at time \emph{t} or not. Time \emph{t} goes from time \emph{0} to \emph{t$_{\text{max}}$} in integer steps, where \emph{t$_{\text{max}}$} is the sum of the execution times of all operations and the shipping time between machines.

The objective function \emph{H(x)} is defined as a cost function (Eq.~\ref{function}), which minimizes the time taken to complete all operations (the makespan). Cost functions are mathematical expressions of optimization problems. For quantum computing, cost functions are quadratic models that have the lowest values (energy) for good solutions to the problems they represent. Each point on a cost function represents a different solution configuration - in this case, each configuration is a particular assignment of the starting times of operations. The goal of the optimization is to minimize the cost of the solution, i.e., the amount of time needed to complete all operations. The mathematical representation of the penalty terms and the makespan objective are added and build the cost function of the \ac{DFJSP} with the format:

\begin{equation}
H(x)=\alpha * f(x)+\beta * g(x)+\gamma * h(x)+k(x)\label{function}
\end{equation}

where: 
\begin{description}
  \item\emph{f(x)}, \emph{g(x)}, and \emph{h(x)}  represent penalty functions for precedence, operation once, and no-overlap constraints, respectively,
  \item\emph{k(x)} represents the objective function, which will minimize the makespan,
  \item${\alpha}$, ${\beta}$, ${\gamma}$  are the Lagrange parameters that emphasize different weights assigned to penalties.
\end{description}

Every time a solution violates one or more constraints, a penalty is applied to that solution (e.g., two operations on a single machine that are scheduled to run at the same time violate the no-overlap constraint, which means that a penalty will be assigned). Positive terms penalize certain solution configurations, while negative ones support them. By adding penalties to terms that break the constraints, the cost of those configurations increases. This reduces the likelihood of generating a suboptimal solution. The importance attributed to each term can be adjusted using the weights ${\alpha}$, ${\beta}$, and ${\gamma}$. The process of adjusting these weights is explained in Section~\ref{sec:penalties}. The weights determine the importance of penalty functions in relation to other terms.

\textbf{Precedence constraint.} The precedence constraint states that a job operation can only start after the previous operation is finished and the lot arrived at the production site of the next operation, if shipping is necessary.

To formulate the penalty, cases where consecutive operations \emph{o} and \emph{o$'$} in a job take place out of order are counted. If the starting time of \emph{o$'$} denoted by \emph{t$^{\prime}_{\text{i,o',m'}}$} is less than the sum of the starting time of \emph{o} denoted by \emph{t$_{\text{i,o,m}}$}, its processing time \emph{p$_{\text{i,o,m}}$} and the shipping time between machines \emph{d$_{\text{m,$m'$ }}$}, then this counts as penalty. The penalty condition can be formulated as follows:

\begin{equation}
t^{\prime}_{i,o',m'}-t_{i,o,m}<p_{i,o,m} + d_{m, m^{\prime}}
\end{equation}

This penalty is the sum of all operations of a job \emph{i} for all the jobs:

\begin{equation}
\begin{split}
      f(x) = &\sum_{\mathclap{i \in J}}\\
      &\hspace{2.0em}\sum_{\mathclap{\substack{o, o^{\prime} \in O_i \\ r_{o}~<~r_{o^{\prime}}}}}\;\\
      &\hspace{4.0em}\sum_{\mathclap{\substack{t^{\prime}_{i,o',m'}-t^{~}_{i,o,m}~<~p_{i,o,m} + d_{m, m^{\prime}} \\ m, m^{\prime} \in M_{i,o} \times M_{i,o^{\prime}}}}}x_{i, o, m, t} \cdot x_{i, o^{\prime}, m^{\prime}, t^{\prime}}
\end{split}
\end{equation}

where:

\begin{description}
  \item[$\sum_{i \in J}:$] sum over all jobs,
  \item[${\sum_{o, o^{\prime} \in O_i}}:$] sum over all combinations of sequential operations,
  \item[${t^{\prime}_{i,o',m'}-t_{i,o,m}<p_{i,o,m} + d_{m, m^{\prime}}}:$] penalty condition,
  \item[${x_{i, o, m, t} \cdot x_{i, o^{\prime}, m^{\prime}, t^{\prime}}}:$] represents the expression that assigns a penalty every time the precedence constraint is violated.
\end{description}

\textbf{Operation-once constraint.} To represent this mathematically, the following function is formulated, which sums values across all possible time steps \emph{t}:

\begin{equation}
\begin{aligned}
g(x)=\sum_{i \in J} \sum_{o \in O_i}\left(\sum_{t_{i,o} \in T} \sum_{m \in M_{i,o}} x_{i, o, m, t} -1\right)^2
\end{aligned}
\end{equation}

where:

\begin{description}
  \item[\emph{${\sum_{i \in J}}:$}] sum over all jobs,
  \item[\emph{${\sum_{o\in O_i}}:$}] sum over all operations,
  \item[\emph{${\sum_{t_{i,o}\in T} \sum_{m\in M_{i,o}}}:$}] sum over all time steps on all machines,
  \item[\emph{${x_{i, o, m, t}}:$}] represents the expression that assigns a penalty every time the precedence constraint is violated,
   \item[\emph{${\sum(...)^2:}$}] squares the results in parentheses to keep the penalty term positive in case an operation is not scheduled.
\end{description}

\textbf{No-overlap constraint.} The penalty function is the sum of pairwise products of binary variables \emph{${x_{\text{i}, \text{o}, \text{m}, \text{t}}}$} and \emph{${x_{\text{i$^{\prime}$}, \text{o$^{\prime}$}, \text{m}, \text{t$^{\prime}$}}}$} that represent starting time of operations on each machine, which must be equal to zero for the configuration to be valid:

\begin{equation}
%\begin{aligned}
h(x)=\sum_{\mathclap{m \in M_{i,o} \cap M_{i^{\prime},o^{\prime}}}}\hspace{2.75em}
\sum_{\mathclap{i,i^{\prime} \in J}}\hspace{2.75em}
\sum_{\mathclap{\substack{o, o^{\prime} \in O_i \times O_i^{\prime}\\
i,i^{\prime}, t,t^{\prime} \in A_m \cup B_m}}}
x_{i,o,m,t} \cdot x_{i^{\prime}, o^{\prime}, m, t^{\prime}} 
%\end{aligned}
\end{equation}

The sets $A_m$ and $B_m$ are defined as follows:

\begin{equation*}
\begin{split}
A_m&=\{(i, i^{\prime}, t, t^{\prime}): i, i^{\prime} \in J, i \neq i^{\prime}, t, t^{\prime} \in T, 0<t^{\prime}-t<p_{i,o,m}\} \\
B_m&=\{(i, i^{\prime}, t, t^{\prime}): i, i^{\prime} \in J, i \neq i^{\prime}, t=t^{\prime}, p_{i,o, m}>0,  p_{i^{\prime},o^{\prime}, m}>0\}
\end{split}
\end{equation*}

The set \emph{A${_m}$} is a condition that forbids operation \emph{o${^{\prime}}$} to start at time \emph{t$^{\prime}$} if the processing of the previous operation \emph{o} has not finished yet (meaning that $\emph{t$^{\prime}$} - \emph{t} < \emph{p$_{\text{i,o,m}}$}$). The set \emph{B${_m}$} forbids the operations to start at the same time. Further details on the penalty function $h(x)$ are:

\begin{description}
  \item[\emph{${\sum_{m \in M_{i,o} \cap M_{i^{\prime},o^{\prime}}}}:$}] sum of each independent machine,
  \item[\emph{${\sum_{i, i^{\prime} \in J}}:$}] sum over jobs,
  \item[\emph{${\sum_{o, o^{\prime} \in O_i \times O_i^{\prime}}}:$}] sum over operations,
  \item[\emph{$x_{i, o, m, t} \cdot x_{i^{\prime}, o^{\prime}, m, t^{\prime}}:$}] product of binary variables, that assigns a penalty every time the no-overlap constraint is violated.
\end{description}

\textbf{Makespan Function.} To formulate the optimization objective of completing each job in the shortest time possible, the following polynomial function is defined, which penalizes the late completion time of operations:

\begin{equation}
\begin{aligned}
k(x)=\sum_{\substack{o \in O_i\\
        m  \in M_{i,o}}}
        x_{i, o, m, t} \cdot (t_{i,o,m} + p_{i,o,m}-P_{i,o})
\end{aligned}
\label{makespan}
\end{equation}

where:

\begin{equation*}
P_{i,o}=\sum_{\substack{r_{o}^{\prime}~<~r_{o}\\ 
        m^{\prime} \in M_{i,o^{\prime}}}}
        {\min}(p_{i,o^{\prime}, m^{\prime}}+d_{m, m^{\prime}})
\end{equation*}

\emph{P$_{\text{i,o}}$} is the minimum predecessor time of the operation \emph{o}, which is the sum of the minimum processing times of the preceding operations and shipping distances between them. 

%
% Subsection: Variable pruning
%
\subsection{Variable pruning}
\label{sec:variable_pruning}

In the previous Section, the binary variable \emph{x$_{\text{i,o,m,t}}$} was defined to represent whether an operation \emph{o} of a job \emph{i} is scheduled to start on a machine \emph{m} at time \emph{t} or not. To reduce the number of variables in the model, a method called variable pruning is used, which removes variables that are not possible in valid solutions. First, an upper bound \emph{t$_{\text{max}}$} of the highest possible makespan is calculated. Without this step, the solver will have to consider all solutions between zero and infinity, which implies a very large solution space that makes it difficult to embed and solve the \ac{QUBO}~\citep{49:venturelli2016}. Instead, the search space is limited by defining the upper bound of a schedule as the sum of the maximum processing time of each operation in all jobs and the maximum shipping time between these operations. Thus, a significant number of variables \emph{x$_{\text{i,o,m,t}}$} can be pruned by applying restrictions on time \emph{t}.

Next, the duration and order of operations within a job are examined. Variables that prevent an operation from executing at a given time, resulting in an invalid schedule, are removed. Specifically, variables are pruned to allow operations to start earlier than the sum of the processing time of its predecessors, or if they allow operations to begin later than the sum of its own processing time and that of the successor operations subtracted from \emph{t$_{\text{max}}$}~\citep{123:Zielewski2020}.   

To set a lower bound of operations in a job, the notion of minimum predecessor time \emph{P$_{\text{i,o}}$} of an operation is used:

\begin{equation*}
P_{i,o}=\sum_{\substack{r_o^{\prime}~<~r_o\\ 
        m^{\prime} \in M_{i,o^{\prime}}}}
        {\min}(p_{i,o^{\prime}, m^{\prime}}+d_{m, m^{\prime}})
\end{equation*}

\emph{P$_{\text{i,o}}$} is the minimum predecessor time of the operation \emph{o}, calculated as the sum of the minimum processing time of the preceding operations and shipping distance between them. A starting time earlier than the minimum predecessor time is invalid; therefore, these binary variables can be eliminated from the model~\citep{103:Schworm2023}.

Additionally, an upper bound for the starting time \emph{t} of operations can be set. This corresponds to the maximum makespan \emph{t$_{\text{max}}$} minus the sum of the processing time of the operation and that of all subsequent operations and shipping time between machines  \emph{S$_{\text{i,o}}$}:

\begin{equation}
t_{max}-p_{i,o,m}-S_{i,o} 
\end{equation}

where:

\begin{equation*}
S_{i,o}=\sum_{\substack{r_o^{\prime}~\geq~r_o\\ 
        m,m^{\prime} \in M_{i,o'}}}
        {\min}(p_{i,o_i^{\prime}, m^{\prime}}+d_{m, m^{\prime}})
\end{equation*}

These variables can also be eliminated from the \ac{QUBO}, since they represent invalid starting times~\citep{49:venturelli2016}. Thus, the starting time of operations \emph{o}  should be in the range $|\emph{P$_{\text{i,o}}$}, \emph{t$_{\text{max}}$} - \emph{S$_{\text{i,o}}$}|$, all the other variables that are not in this range are removed from the model to decrease the size of the computational problem and computation time~\citep{123:Zielewski2020}.

To further reduce the number of binary variables, the starting times of simultaneous operations on machines are also removed~\citep{68:Denkena2021}.

%
% Subsection: Penalties
%
\subsection{Lagrange parameters}
\label{sec:penalties}

In the previous sections, the constraints have been formulated as binary polynomials in a Hamilton formulation. The binary polynomials \emph{f(x)}, \emph{g(x)}, \emph{h(x)} are multiplied with non-negative scalar weights ${\alpha, \beta, \gamma}$ also known as Lagrange parameters (Eq.~\ref{function}), which determine the impact of the respective polynomial~\citep{103:Schworm2023}. Weights are assigned to each constraint depending on the relative strengths of the objective function and the constraints. The weights must be set high enough so that a valid solution with no broken constraints is returned. On the other hand, values that are too high will result in solutions that are valid, but far from optimal\footnote{Problem Formulation Guide [White Paper]:~\url{https://www.dwavesys.com/media/bu0lh5ee/problem-formulation-guide-2022-01-10.pdf}}. 

To find the Lagrange parameter ${\beta}$ for the operation-once constraint, the case of \emph{H(x)} is investigated that consists of the constraint \emph{g(x)} and the makespan objective function:

%%% earlier version, with difference to g(x)
%%% \begin{gather*}
%%% H(x) =  \beta * \left(\sum_{i \in J} \sum_{o_i \in O_i}\left(\sum_{t_{o_i} \in T} \sum_{m \in M_{o_i}} x_{i, o, m, t}-1\right)^2\right)\\
%%% +\sum_{\substack{o \in O_i\\
%%%         m  \in M_{i,o}}}
%%%         x_{i, o, m, t} \cdot(t_{i,o,m}+p_{i,o,m}-P_{i,o})\\
%%% \label{eq:2}
%%% \end{gather*}

\begin{equation}
\begin{split}
    H(x) = \beta & * \left(\sum_{i \in J} \sum_{o \in O_i}\left(\sum_{t_{i,o} \in T} \sum_{m \in M_{i,o}} x_{i, o, m, t}-1\right)^2\right) \\
    & +\sum_{\substack{o \in O_i\\
            m  \in M_{i,o}}}
            x_{i, o, m, t} \cdot(t_{i,o,m}+p_{i,o,m}-P_{i,o})
\label{eq:2}
\end{split}
\end{equation}

For example, for job \emph{j${_1}$} in Tbl.~\ref{tab:lagrange}, operation \emph{o${_1}$} is starting on machine \emph{m$_{\text{1}}$} at time \emph{t$_{\text{0}}$} and has a processing time \emph{p$_{\text{1,1,1}}=1$} (this technique is applied in the same way for larger problem instances with more jobs and operations). 

Thus, the sum of processing times of previous operations \emph{${P_{\text{1,2}}}$} of \emph{o${_2}$} is equal to 1. Further, \emph{o${_2}$} of the same job is starting at different times \emph{t} on machine \emph{m$_{\text{3}}$}. Shipping time between machines \emph{d$_{\text{1,3}}$} equals to zero. Both binary variables \emph{x$_{\text{1,1,1,0}}$} and \emph{x$_{\text{1,2,3,1}}$} are equal to 1. This can be expressed by a \ac{QUBO}. For \emph{$\beta=1$}, the value and thus, the energy of cost function \emph{H(x)} equals to 3:

\begin{align*}
 &= \beta \cdot ((x_{1, 1, 1, 0}-1)^2+(x_{1, 2, 3, 1}-1)^2) +(x_{1, 1, 1, 0}+2\cdot x_{1, 2, 3, 1}) \\
&=1 \cdot ((1-1)^2+(1-1)^2)+(1+2 \cdot 1)=3
\label{eq:4}
\end{align*}

This is the sum of the processing times of the two operations \emph{o${_1}$} and \emph{o${_2}$} that represent the makespan objective function (second term in equation \emph{H(x)}). Since the start-once constraint is not violated, the first term in equation \emph{H(x)} is equal to zero. This is also the minimum energy level of a valid solution for this specific problem instance. Any violation of the constraints will add penalties to the cost function that will increase the energy of the system.

A specific case, however, should be considered. Let us assume that operation \emph{o${_2}$} is not scheduled to start at all. This will result in a cost function \emph{H(x)} that equals to 2:

\begin{gather*}
H(x)=1\cdot((1-1)^2+(0-1)^2)+(1+0)=2
\end{gather*}

Although the returned solution is invalid, the energy of the system equals $2$ in this case, which is less than the minimum valid energy of $3$. Tbl.~\ref{tab:lagrange} summarizes the results for function \emph{H(x)} for \emph{$\beta=1$}.

%
% Table: Lagrange
%
\begin{table}[ht]
%\setlength{\tabcolsep}{3pt}
%\begin{tabular}{w{c}{50pt}w{c}{55pt}w{c}{55pt}w{c}{55pt}}
\begin{center}
\begin{tabular}{c c c c}
\toprule
\emph{t} & \emph{${x_{\text{1}, \text{1}, \text{1}, \text{t}}}$} & \emph{${x_{\text{1}, \text{2}, \text{3}, \text{1}}}$} & \emph{${x_{\text{1}, \text{2}, \text{3}, \text{t}}}$} \\
\midrule
0 & 1 & 0 & 0 \\
1 & 0 & 1 & 0 \\
2 & 0 & 0 & 0 \\
3 & 0 & 0 & 0 \\
\emph{${H(x)}$} & - & 3 & 2 \\
\bottomrule
\end{tabular}
\end{center}
\caption{Lagrange parameters in cost function.}
\label{tab:lagrange}
\end{table}

To compensate for this effect, a Lagrange parameter is applied that increases the weight of the penalty term. In \ac{DFJSP}, valid solutions with higher energy are preferred to invalid solutions with low energy but violated constraints. The value of the weight parameter depends on the problem instance. A value that is too low penalizes invalid configurations of variables insufficiently, resulting in energy values that might be lower than the minimum energy of the best valid solution. Values that are too high would return valid configurations -- possible with higher energy than the optimum. The value of the Lagrange parameter is set based on the maximum makespan of the jobs ( \emph{t$_{\text{max}}$} in Section~\ref{sec:variable_pruning}). The maximum energy value, which corresponds to the maximum makespan \emph{t$_{\text{max}}$}, is equal to 10 (this value is calculated for each problem instance based on the input job dictionary). The penalty is set to \emph{$\beta=t_{\text{max}}$}, meaning the maximum energy possible for a valid configuration for this problem instance, which results in the following function \emph{H(x)}:

\begin{gather*}
H(x)=(1+0)+10 \cdot ((1-1)^2+(0-1)^2)=11
\end{gather*}

The energy of the function is equal to $11$, which is above the maximum energy possible for valid solutions of $10$. This ensures that no invalid solution will have lower energy than any of the valid solutions. The technique demonstrated in this simplified example can be applied similarly to problems with more jobs and operations. The resulting \ac{QUBO} will have more variables.  

The same method is used to calculate the weights for other constraints as well. Generally, the Lagrange parameters are set individually for each constraint, depending on the magnitude of the objective function to be minimized and its relative strength to the constraints. In this case, to make sure that the cost function \emph{H(x)} has the lowest energy for valid solutions, the weights ${\alpha, \beta, \gamma}$ of penalty functions \emph{f(x)}, \emph{g(x)} and \emph{f(x)} are all set equal to \emph{t$_{\text{max}}$}, which is the maximum energy value of all valid solutions.

\section{Experiments}\label{sec:exp}

In this Section, we implement the model and conduct experiments on a D-Wave quantum annealer (specifically, the \mbox{\textit{Advantage System 4.1}}). As is common practice, the results of the experiment on a quantum annealer are compared with those obtained from SA. Firstly, the problem size (see~\ref{sec:dataset}) and solver parameters (see~\ref{sec:qpu_parameters}) are defined. The \ac{QUBO} model is constructed using the PyQUBO Python library, which provides a high-level interface for formulating combinatorial optimization problems in \ac{QUBO} form and supports solving them using various solvers, including the D-Wave quantum annealer~\citep{121:zaman2021pyqubo}. We then present the results of the experiments in Subchapter~\ref{sec:chapter04:listen} and discuss them in Subchapter~\ref{sec:discussion}.

\subsection{Problem size and dataset}
\label{sec:dataset}

To analyze the performance of the D-Wave quantum annealer for the \ac{DFJSP}, 6 problem instances are defined in such a way that the size of the smallest problem instance is 50 and the size of the largest one is 250. The instances in between are defined in even intervals of 50 (e.g., 100, 150, etc\footnote{Problem size is rounded for some instances (e.g. instance size 99 is rounded to 100).}). The size of the problem instances refers to the number of binary variables \emph{x$_{i, o, m, t}$} necessary to represent the problem in \ac{QUBO} form. The maximum size of the problem instances is dictated by the physical properties of the D-Wave \ac{QPU}. The D-Wave \mbox{\textit{Advantage System 4.1}} has 5627 qubits. However, not all qubits are interconnected. When embedding the model variables onto the physical hardware graph of the \ac{QPU}, in some cases, to represent a single logical variable set of physical qubits called \emph{chains} are necessary, with couplings between physical qubits capable of realizing the correct interaction between logical variables~\cite{149:Yarkoni2022}. This leads to embedding patterns that need more physical qubits than there are variables in the original problem. 

The purpose of the experiment is to test the capacity and performance of the D-Wave quantum annealer for the \ac{DFJSP} up to the largest problem that can be embedded into the \ac{QPU}. Tbl.~\ref{tab:var_qubits} summarizes the instances solved by \ac{QA}, the number of qubits needed to embed the model onto the physical \ac{QPU}, and the length of the longest chain in the embedding.

\begin{table}[ht]
\begin{center}
%\begin{tabular}{w{r}{70pt}w{r}{75pt}w{r}{70pt}}
\begin{tabular}{r r r}
\toprule
Problem size& Number of qubits& Max chain \\
\midrule
$50$& 
$223$& 
$7$ \\
$100$& 
$977$& 
$18$ \\
$150$& 
$2115$& 
$25$ \\
$200$&
$3156$& 
$28$\\ 
$250$&
$4067$& 
$32$ \\
$300$&
%no embedding found&
-&
- \\
\bottomrule
\end{tabular}
\end{center}
\caption{Logical problem variables and number of qubits needed for embedding.}
\label{tab:var_qubits}
\end{table}

Smaller problem instances require a lower number of qubits for embedding. It is also true that the length of the chains increases with problem size, meaning that with an increasing number of variables, it becomes more challenging to find embedding schemas. Therefore, longer chains are required to connect the variables through couplers. In this model, the largest problem instance that could be embedded was of size 250, which required more than 4000 qubits for embedding. For 300 variables, no embedding could be found.

The problem instances are formulated for 2 jobs, \emph{j$_{\text{1}}$} and \emph{j$_{\text{2}}$}, each job containing a minimum of 2 operations. Each operation can be processed on one machine or a set of two machines, with different processing times on each machine. To include shipping distance in the model, the sets of machines available for each operation include machines from different factories. The consideration of the shipping distance is essential for the processing of \textbf{Contribution 1}. The number of operations in each job, the available machines, and the processing time on each machine determine the number of binary variables \emph{x$_{i, o, m, t}$} necessary to represent the problem in \ac{QUBO} form. 

%
% Subsection: QPU configuration
%
\subsection{Solver parameters}
\label{sec:qpu_parameters}

There are numerous solver parameters\footnote{\url{https://docs.dwavequantum.com/en/latest/industrial_optimization/index_properties_parameters.html}} that can be adjusted when submitting problems to solvers. In this experiment, the default settings are kept for most of them, including the default annealing time of 20${\mu}$s. The parameters that are adjusted in the experiment are the number of reads and the chain strength. 

\textbf{Number of reads.} In theory, very long annealing times should keep the system in a low energy state, in practice, noise due to hardware imperfections will push the state to a higher energy~\cite{102:Carugno2022}. Therefore, short annealing times are preferred in order to take many samples from the energy distribution of the problem. The appropriate number of samples depends on the problem itself and its energy distribution, and is usually between 10 and 10,000 samples~\cite{102:Carugno2022}. In this experiment, the number of reads ('num{\_}reads') is set to 1000. Generally, increasing the number of reads can increase the probability of obtaining better solutions, but improvements diminish due to some problem-dependent factors. For this model, increasing the number of reads significantly above 1000 does not appear to be advantageous. Sampling the problem instance with 200 variables 3000 times produced a solution with an energy of 34 and 2 violated constraints, which is worse than the energy of the best solution of 24 and 1 violated constraint obtained with 1000 reads. This is consistent with the results obtained by~\citeauthor{102:Carugno2022}~\citep{102:Carugno2022}, who found that increasing the number of reads from 1000 to 10000 did not have any beneficial effects. 

\begin{figure}[ht]
    \centering
    \includegraphics[width=0.8\textwidth]{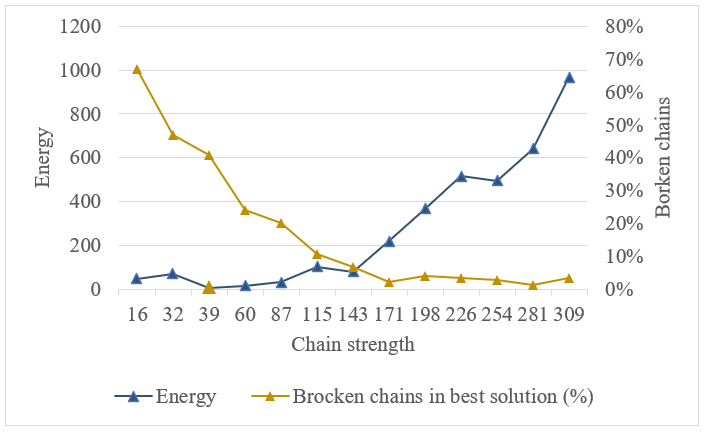}
    \caption{Chain break analysis for problem instance of size 150 (Source: authors).}
    \label{fig:chain_strength_broken}
\end{figure}

\textbf{Chain strength.} For a solution to be consistent, the chain of qubits has to act as a single qubit, meaning that all the qubits in the chain must have the same values after annealing~\cite{102:Carugno2022}. Broken chains show different values of the qubits at the end of the annealing process. Chains must be set strong enough to ensure that the qubits of the chain show the same value compared to other qubits. If the chain's strength is too small, then the qubits will show different values. If the chain strength parameter is too high, it will impede the qubits in the chain from changing their state, causing the chain to behave like a separate entity that does not interact with other terms of the function. 

We follow the D-Wave recommendation\footnote{\url{https://docs.dwavequantum.com/en/latest/quantum_research/embedding_guidance.html}} to determine the chain strength for each problem instance and start with two utility functions to estimate the chain strength: 

\begin{enumerate}
\item \textit{scaled}
\item \textit{uniform{\_}torque{\_}compensation}
\end{enumerate}

The first function returns a chain strength that is scaled to the problem bias range. The second function calculates a chain strength that attempts to compensate for the torque that would break the chain. Both methods should be viewed as a starting point for estimating the optimal chain strength, rather than a rule, since this parameter is problem-dependent. In the next step, the chain strength parameter found in step 1 should be doubled or tripled to check if this change has an impact on the percentage of broken chains.

To find the best chain strength for the model, the model is run on the \ac{QPU} for each problem instance, and the resulting energy and the percentage of broken chains in the best solution are analyzed. The intention is to find the lowest energy possible, while also having a low rate of broken chains to ensure a consistent solution. Fig.~\ref{fig:chain_strength_broken} shows the results of the measurements for the problem instance of size 150.

\begin{figure}[ht]
    \centering
    \includegraphics[width=0.8\textwidth]{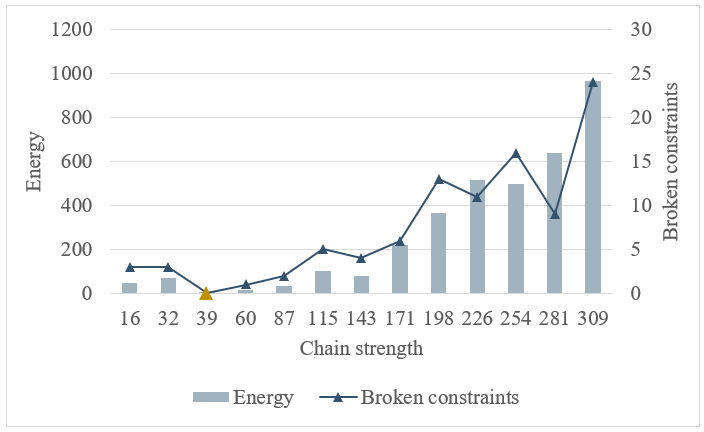}
    \caption{Chain strength impact on solution quality (Source: authors).}
    \label{fig:chain_strength_solution}
\end{figure}

The utility function \emph{scaled} returns a chain strength of 32 with an energy of 69; the function \emph{uniform{\_}torque} returns a value of 309 with an energy of 966. To understand the relationship between these parameters, the range between these two chain strengths is divided into 11 intervals, and the energy and the percentage of broken chains in the best solutions for each of these points are measured. Additionally, the smallest chain strength value is divided by 2 to check whether a smaller parameter could improve the results. The results in the graph show that the lowest energy achieved corresponds to the next point in the interval from the point obtained with the \emph{scaled} function.
%(the method to find this level is referred to as \emph{manual} further)
However, the percentage of broken chains corresponding to this value is 41{\%}, which is high compared to values in the graph for higher chain strength values. To determine whether to accept a chain strength parameter of 39, an additional analysis of the solution's consistency is necessary. Fig.~\ref{fig:chain_strength_solution} shows that, despite a higher number of broken chains, the embedding with a chain strength of 39 delivers a consistent solution without broken constraints. As the chain strength parameter increases, the number of broken constraints also increases.

This procedure is repeated in the experiment for all problem instances to calculate the appropriate parameter for the chain strength. For all instances, the parameter that delivered the best solution with the lowest energy is one interval above the parameter estimated by the \emph{scaled} function. Since the intention is to test the extent to which the \ac{QPU} can handle large problems, this method is applied to problem instances of sizes 250 and 300. For problems of size 50 to 150, the solver can find consistent solutions with no broken constraints. For problem instances of size 200 and 250, the solutions with the lowest energy found had 1 and 2 broken constraints, respectively. For the instance of size 300, no embedding could be found.

\textbf{SA parameters.} To solve problems with \ac{SA}, the D-Wave Sampler class \textit{SimulatedAnnealingSampler} of the D-Wave Ocean SDK is used. The sampler utilizes a classical heuristic \ac{SA} algorithm that runs locally on the CPU\footnote{Hardware configuration: 2x Intel Xeon Gold 6326 CPU 16Core, 2.9GHz 24M 185W Socket 4189, 4 x Samsung 64GB DDR4 ECC RDIMM 
3200Mhz total 256GB RAM, 4 x HDD WD Gold Datacenter 1TB SATA, 1x SSD Samsung PM883 480GB SATA, 2 x 1 / 10 GB Ethernet 10 BaseT RJ-45 onboard.}. For \ac{SA}, the \ac{CPU} time is measured, which is the amount of time that the \ac{CPU} has spent solving the problem. The \ac{CPU} time is compared to \ac{QPU} access time. 

To compare \ac{QA} and \ac{SA}, the number of iterations needed and the ability to find a consistent solution are analyzed. To have comparable results, the number of reads for \ac{SA} is set to 1000, like the parameter for \ac{QA}. The parameter number of sweeps  (\emph{num{\_}sweeps}) is set to 3, which is the smallest value for this parameter that returned solutions with no broken constraints. As with \ac{QA}, the \ac{SA} algorithm is run for all the instances of the problem to determine whether the system is able to find valid solutions. Additionally, problem instances of size 350 and 400 are included in the analysis.

\subsection{Numerical Results}\label{sec:chapter04:listen}
To provide the \textbf{Contributions 2 and 3} formulated in Section~\ref{sec:intro}, the results obtained from the experiment are analyzed from both a qualitative and a quantitative perspective. The first answers the question of how good the returned solution is in terms of system energy and makespan. The second compares \ac{QPU} access time and \ac{CPU} time of \ac{SA} to assess whether there is a speed advantage in using the quantum annealer to solve the \ac{DFJSP} with distributed operations. The hardware characteristics influence the results; different hardware may yield different results.

\begin{figure}[ht]
    \centering
    \includegraphics[width=0.8\textwidth]{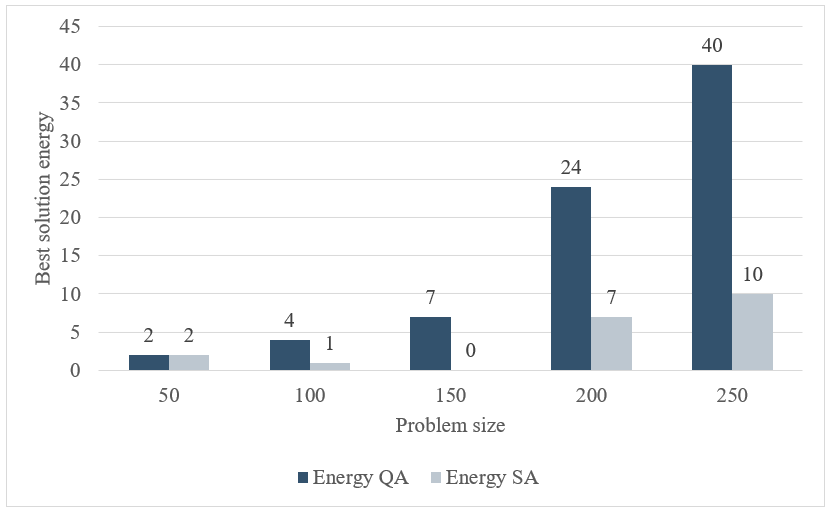}
    \caption{Energy of best solutions obtained with QA and SA (Source: authors).}
    \label{fig:energy}
\end{figure}

\textbf{System energy and makespan.} The goal of the experiment is to find consistent solutions in the shortest time possible. When sampling from \ac{QA} or \ac{SA}, the returned sample set contains 1000 samples, in accordance with the defined number of reads. The sample with the lowest energy in the sample set is selected as the best sample result for the respective problem instance. This is not always the solution corresponding to the global minimum, which is represented by the lowest energy possible for the \ac{QUBO}. For this use case, however, finding the global minimum is not crucial. A solution is satisfactory if it is consistent and represents a valid schedule plan that can be obtained in a reasonable amount of time. Further optimization steps can then be implemented to minimize the energy of the system and the makespan. Fig.~\ref{fig:energy} shows a comparison of best solution energies found with \ac{QA} and \ac{SA} for instances of different sizes.

For the smallest problem instance, both \ac{QA} and \ac{SA} found a low energy of 2 and returned valid solutions. For instances of size 100 and 150, both methods delivered consistent solutions, but \ac{SA} performed better in terms of energy. For problem instances of size 200 and 250, the best solution returned by the quantum annealer had 1 and 2 broken constraints, respectively. This explains the increasing energy values for these problems. This observation is related to the percentage of broken chains, which increases with problem size and deteriorates solution quality as the number of variables in the problem grows. These results are consistent with results found in other papers.~\citeauthor{102:Carugno2022}~\cite{102:Carugno2022} found that for \ac{JSSP} instances with an unknown optimal solution, \ac{QA} is not able to find the best solution that corresponds to the global minimum. 

\begin{figure}[ht]
    \centering
    \includegraphics[width=0.8\textwidth]{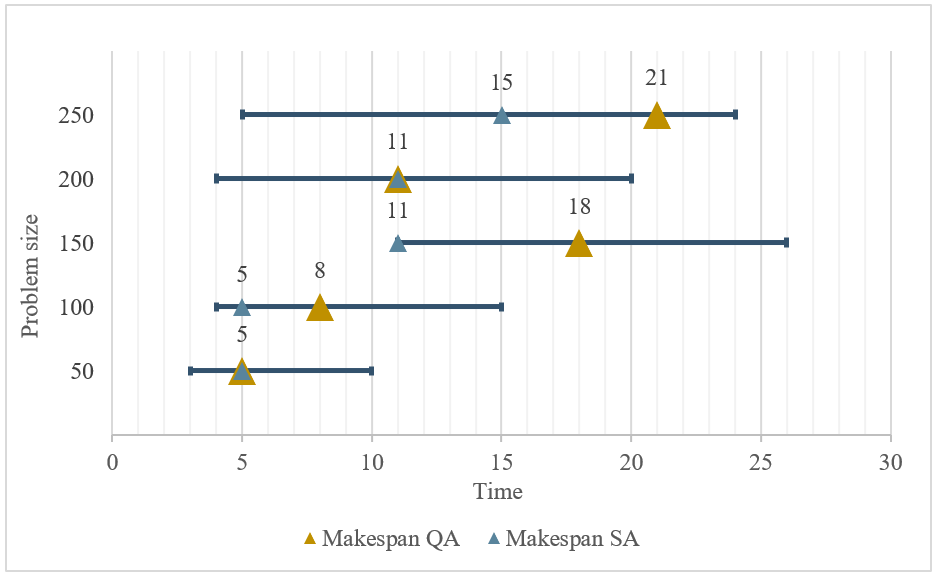}
    \caption{Comparison of Makespan values obtained by \ac{QA} and \ac{SA} (Source: authors).}
    \label{fig:makespan}
\end{figure}

To calculate the makespan in the \emph{makespan function} (Eq.~\ref{makespan}), the minimum ending time and the maximum ending time of all operations are used. These terms represent the minimum and the maximum makespans possible and are the lower and the upper bound of the makespan range, respectively. In terms of makespan, \ac{QA} returns solutions in the lower half of the possible makespan range for all problem sizes, except for size 250, which is in the upper half of possible solutions (Fig.~\ref{fig:makespan}).

To summarize, \ac{QA} can find consistent solutions for problem instances up to 150 variables. For the instance of size 200, the makespan found is still in the first half of solutions, but 1 constraint is violated. For size 250, the makespan is slightly below the maximum possible value, and 2 constraints are violated. Current \ac{QA} technology is subject to hardware and software limitations; however, with the growing number of qubits and increasing connectivity between them, problems with a higher number of variables can potentially be embedded into future \ac{QPU}s. Additionally, improving \ac{QPU} stability promises to deliver solutions of better quality and with lower energy of the system.

Next, \ac{QPU} access time and \ac{CPU} time of \ac{SA} for the \ac{DFJSP} with distributed operations are compared. 

\begin{figure}[ht]
   \centering
   \includegraphics[width=0.8\textwidth]{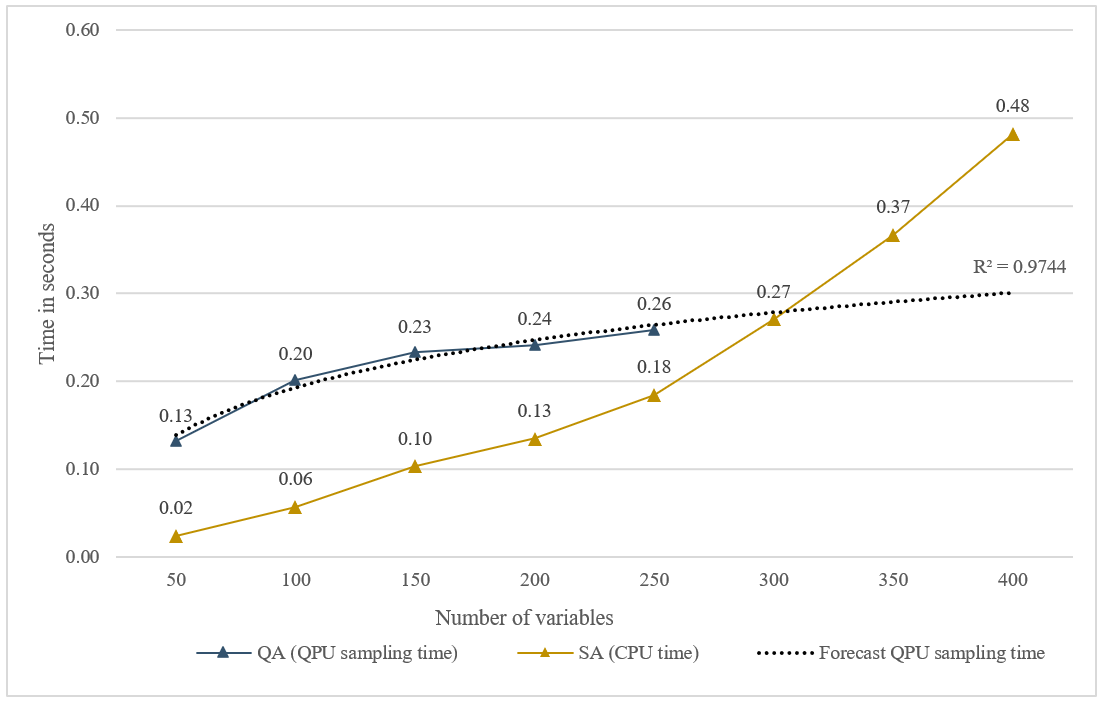}   \caption{Calculation time with QA and SA (Source: authors).}   \label{fig:runtime}
\end{figure}

\textbf{Calculation time.} To determine the \ac{QPU} access time, the instances from size 50 to 250 are solved with the quantum annealer. Based on these results, a trendline is derived for instances of size 300 to 400. In this case, since the rate of change in the data decreases and levels off, a logarithmic trendline best describes the data. The R-squared value of 0.9767 is a strong indicator that the logarithmic trendline accurately represents the data. With \ac{SA}, it was possible to solve problem instances of all sizes, up to the 400-variable problem instance. Fig.~\ref{fig:runtime} shows the results that were obtained by both methods. 

Although \ac{SA} performs better for small problem sizes, the \ac{CPU} calculation time of \ac{SA} increases exponentially with the increasing problem size. On the other hand, due to hardware limitations, the \mbox{\textit{Advantage System 4.1}} quantum annealer is capable of embedding problems up to a maximum size of 250 variables. However, its advantage is that the calculation time increases at a decreasing rate with problem size. For business problems with a very large number of variables, specific to the \ac{DFJSP}, a calculation time that scales logarithmically with problem size is a significant benefit, compared to the exponential growth of \ac{CPU} time.

\subsection{Discussion}
\label{sec:discussion}

In this paper, a step-by-step practical implementation of the \ac{DFJSP} with distributed operations and its evaluation using D-Wave quantum annealer, based on a real use case from the wool textile industry, is presented. First, the problem is described mathematically as a constraint satisfaction problem. Then, the process of reformulating the mathematical problem in \ac{QUBO} format is discussed in detail, as well as its embedding into the D-Wave quantum annealer \ac{QPU} and the configuration of \ac{QPU} solver parameters. Finally, solutions are sampled from the \ac{QPU} to find low-energy states of the cost function. While formulating the \ac{QUBO}, particular attention is dedicated to the calculation of the appropriate Lagrange parameters (Subsection~\ref{sec:penalties}), which increase the weights of the penalty terms. Another aspect that is carefully investigated while configuring the experiment is the chain strength parameter (Subsection~\ref{sec:qpu_parameters}). Both aspects make up \textbf{Contribution 2}.

Another aim of the paper was to investigate whether \ac{QA} can provide a speed advantage for the \ac{DFJSP} with operations specific to the wool textile industry. To answer this question, both qualitative and quantitative analyses are conducted. Current \ac{QA} technology enables the use of more qubits than classical gate model quantum computers. However, there are still limitations imposed by hardware and software imperfections. With this in mind, the test challenged to the limit the ability of \ac{QA} to deliver consistent solutions for large problem instances in the shortest time possible. To conduct the experiment, problem instances with sizes ranging from 50 to 250 are solved in increments of 50 variables using \ac{QA} and \ac{SA}. Additionally, problems with up to 400 variables are solved only with \ac{SA} to increase the size range and determine the calculation time trend. The missing data for \ac{QA} is extrapolated as a logarithmic trendline. 

The qualitative analysis of the solutions returned indicates that for problem instances with up to 150 variables, both \ac{QA} and \ac{SA}, with technical characteristics as used in this research, deliver consistent solutions, and thus valid production schedules. However, \ac{SA} performs slightly better in terms of returned energy. With an increasing number of variables, hardware imperfections become more visible, manifesting as a higher number of broken chains in the returned solutions, and consequently, in solutions with higher energy due to broken constraints. For the problem of size 200, the best solution returned by \ac{QA} has 1 broken constraint; for the instance of 250 variables, 2 constraints were violated. Due to a higher usage of physical qubits needed compared to the number of variables in the original \ac{QUBO}, the maximum problem instance that could be embedded onto the \ac{QPU} was of 250 variables. For the problem of size 300, no embedding could be found. In terms of makespan, \ac{QA} finds solutions for almost all problem sizes in the lower half of the possible makespan range, except for size 250, which is in the
upper half of the possible solutions. The makespan determined by \ac{SA} is comparable, although for a few problem instances \ac{SA} can find solutions that are closer to the global minimum. All solutions calculated by \ac{SA} are consistent, even for the largest instances in the test. 

To tackle the \textbf{Contribution 3}, \ac{QPU} access time and \ac{SA} \ac{CPU} time necessary to calculate the solutions to the problems were measured. The results reveal that \ac{QA} requires more time than \ac{SA} for smaller problem instances, but the time needed increases at a decreasing rate as the problem size grows. This is not the case for \ac{SA}, whose \ac{CPU} time grows exponentially with the increasing number of variables. The \ac{CPU} timeline and the trendline of the \ac{CPU} access time are slightly above the value of 300 variables. If this trend persists with the new \ac{QA} hardware generation that is to come, every problem with a size larger than this value could be solved faster than with \ac{SA}. Thus, the question of the speed advantage can be answered positively: \ac{QA} can potentially result in a speed advantage for the \ac{DFJSP} with distributed operations. 

Even if the \ac{QA} technology currently faces limitations, the new generation of quantum annealers with improved hardware promises to accelerate the solution of complex combinatorial optimization problems, such as the \ac{DFJSP}. The measurements show that, given the same experimental parameters and improved hardware, \ac{QA} can provide a speed-up advantage over \ac{SA} starting with problem instances that are slightly larger than 300 variables. To test this assumption, new tests must be performed for larger problem instances, for which only the trend could be extrapolated at the moment, as soon as the latest generation of quantum annealers becomes available. More qubits and connections between them, as well as more stable systems, will enable the embedding of larger problems and the discovery of energies closer to the global minimum. Additionally, research on tuning and configuration of \ac{QA} parameters is another necessity that will enable us to minimize energy and find the optimal solution. Determining the strength of the qubit chains is a laborious task that is not only model-dependent but also instance-dependent and has a significant impact on the quality of the obtained solutions.
\section{Conclusion}\label{sec:conclusion}

Many modern manufacturing enterprises have evolved from a single to a multi-factory production environment, as is the case with big production companies in the wool textile industry. Planning systems changed from classical \ac{JSSP} to \ac{DFJSP}, with companies facing the challenge of increased frequency and complexity of production planning. Classical heuristics deliver valid results on small problem instances but are unable to find solutions for problem sizes relevant to practice. Therefore, methods are required that yield high-quality outcomes while ensuring reasonable computation times. In this paper, the concept of \ac{QA}-based optimization for the \ac{DFJSP} with geographically distributed operations is investigated to answer the question of whether \ac{QA} can provide a speed advantage for the \ac{DFJSP} with distributed operations specific to the wool textile industry. 

To do this, we show a step-by-step implementation of the workflow for solving the \ac{DFJSP} with \ac{QA} (\textbf{Contribution 1}). Special attention is devoted to topics that do not have a straightforward approach but a significant impact on the solution quality, such as calculating the Lagrange parameters or determining the chain strength. The results show that up to a certain problem size, \ac{QA} delivers consistent solutions that are not the global minimum solution in the solution space but yield a makespan in the first half of the allowed valid solutions (\textbf{Contribution 2}). For all instances, \ac{SA} delivers comparable or slightly better results, which are consistent with other research. In terms of calculation time, \ac{QA} needs more \ac{QPU} time to find solutions for smaller problem instances in the test, but \ac{QPU} access time increases at a declining rate with the growing problem size. In contrast, the \ac{SA} \ac{CPU} time develops exponentially for problems with a larger number of variables. The results of the experiment show that, starting with a certain problem size, \ac{QA} could potentially offer a speed-up advantage, given that expected improvements in hardware and software are anticipated (\textbf{Contribution 3}). 

One of the future tasks is to solve larger problems on the next generation of quantum annealers to confirm the results obtained in this experiment. Other topics open for future research are \ac{QPU} parameter tuning and configuration. There is also a need for additional research on embedding techniques that can minimize the length of the chains and enable the embedding of more \ac{QUBO} variables into physical qubits, thereby solving problems of greater size.

\bibliographystyle{plainnat}
\bibliography{references}

\end{document}